\documentclass[letterpaper,twocolumn,aps,pra,amssymb,amsfonts,superscriptaddress,showpacs]{revtex4}

\usepackage{verbatim}
\usepackage{amsmath}
\usepackage{graphicx}
\makeatletter
\@ifundefined{textcolor}{}
{%
 \definecolor{BLACK}{gray}{0}
 \definecolor{WHITE}{gray}{1}
 \definecolor{RED}{rgb}{1,0,0}
 \definecolor{GREEN}{rgb}{0,1,0}
 \definecolor{BLUE}{rgb}{0,0,1}
 \definecolor{CYAN}{cmyk}{1,0,0,0}
 \definecolor{MAGENTA}{cmyk}{0,1,0,0}
 \definecolor{YELLOW}{cmyk}{0,0,1,0}
 }

\@ifundefined{definecolor}
 {\@ifundefined{definecolor}
 {\@ifundefined{definecolor}
 {\usepackage{color}}{} }{}  }{}

\usepackage[
colorlinks=true,linkcolor=red,citecolor=green,urlcolor=blue,
pdfstartview=FitH,
bookmarksopen=true,bookmarksopenlevel=0,
plainpages=false,pdfpagelabels,
pagebackref=false,
pdftoolbar=false]{hyperref}
\usepackage{lmodern}

\begin{document}

\title{Limits on Preserving Quantum Coherence using Multi-Pulse Control}

\author{Kaveh Khodjasteh}
\affiliation{\mbox{Department of Physics and Astronomy, Dartmouth
College, 6127 Wilder Laboratory, Hanover, NH 03755, USA}}

\author{Tam\'as Erd\'elyi}
\affiliation{\mbox{Department of Mathematics, Texas A\&M University,
College Station, TX 77843, USA}}

\author{Lorenza Viola}
\affiliation{\mbox{Department of Physics and Astronomy, Dartmouth
College, 6127 Wilder Laboratory, Hanover, NH 03755, USA}}

\begin{abstract}
We explore the physical limits of pulsed dynamical decoupling methods
for decoherence control as determined by finite timing resources. By
focusing on a decohering qubit controlled by arbitrary sequences of
$\pi$-pulses, we establish a non-perturbative quantitative upper bound
to the achievable coherence for specified maximum pulsing rate and
noise spectral bandwidth.  We introduce numerically optimized control
``bandwidth-adapted'' sequences that saturate the performance bound,
and show how they outperform existing sequences in a realistic
excitonic-qubit system where timing constraints are significant.  As a
byproduct, our analysis reinforces the impossibility of
fault-tolerance accuracy thresholds for generic open quantum systems
under purely reversible error control.
\end{abstract}

\pacs{03.67.Pp, 03.67.Lx, 03.65.Yz, 07.05.Dz}
\date{\today}
\maketitle

Building on the discovery of spin-echo and multiple-pulse techniques
in nuclear magnetic resonance \cite{NMR}, dynamical decoupling (DD)
methods for open quantum systems \cite{ViolaDD} have become a
versatile tool for decoherence control in quantum engineering and
fault-tolerant quantum computation.  DD involves ``open loop''
(feedback-free) quantum control based on the application of a
time-dependent Hamiltonian which, in the simplest setting, effects a
pre-determined sequence of unitary operations (pulses) drawn from a
basic repertoire.  Physically, DD relies on the ability to access
control time scales that are short relative to the correlation time
scale of the interaction to be removed.  The reduction in decoherence
is achieved {\em perturbatively}, by ensuring that sufficiently high
orders of the error-inducing Hamiltonian are removed.  Recently, a
number of increasingly powerful pulsed DD schemes have been proposed
and validated in the laboratory.  Uhrig DD (UDD) sequences
\cite{Uhrig2007}, for instance, perturbatively cancel pure dephasing
in a single qubit up to an arbitrarily high order $n$ while using a
minimal number ($n$) of pulses, paving the way to further optimization
for given sequence duration \cite{Biercuk2009,Uys09} and/or specific
noise environments \cite{noise_adapted}, to nearly-optimal protocols
for generic single-qubit decoherence \cite{WestFongLidar2010}.
Experimentally, UDD has been employed to prolong coherence time in
systems ranging from trapped ions \cite{Biercuk2009,Uys09,Steane} and
atomic ensembles \cite{SagiAlmogDavidson2010} to spin-based devices
\cite{Liu2009nat}, and to enhance contrast in magnetic resonance
imaging of tissue \cite{Warren09jcp}.

In a realistic DD setting, the achievable performance is inevitably
influenced by errors due to limited control as well as deviations from
the intended decoherence model. Since it is conceivable that both
model uncertainty and pulse non-idealities can be largely removed by
more accurate system identification and control design, some of these
limitations may be regarded as non-fundamental in
nature. Composite-pulse \cite{Ken} and pulse-shaping \cite{shape}
techinques can be used, for instance, to cancel to high accuracy the
effects of both systematic control errors and finite-width
corrections.  We argue, however, that even in a situation where pulses
may be assumed perfect and instantaneous, an ultimate constraint is
implied by the fact that the rate at which control operations are
effected is necessarily finite -- as determined by a ``minimum
switching time'' $\tau_{\text{min}}$ for the available control
modulation.
Our goal in what follows is to rigorously quantify the performance
limits to preserving coherence using DD as arising from the sole
constraint of {\em finite timing resources}.

We focus on the paradigmatic case of a single qubit undergoing pure
dephasing due to either a quantum bosonic bath at equilibrium or
classical (Gaussian) noise, and controlled through a sequence of
instantaneous $\pi$ pulses.  While representing an adequate
idealization of realistic decoherence control settings
\cite{Biercuk2009,Uys09,Steane,SagiAlmogDavidson2010,Warren09jcp},
this problem is exactly solvable analytically
\cite{ViolaDD,Uhrig2007}, enabling rigorous conclusions to be
established.  Our first result is a {\em non-perturbative} lower bound
for the minimum decoherence error achievable by {\em any} DD sequence
subject to a timing constraint $\tau_{\text{min}}$, for noise spectra
characterized by a finite spectral bandwidth $\omega_c$.  Secondly, we
show how to generate ``bandwidth-adapted'' DD sequences that achieve
optimum performance over a desired storage time while respecting the
pulse-rate constraint, and demonstrate their advantages in a realistic
excitonic qubit.  Conceptually, our analysis highlights connections
between DD theory and complex analysis of polynomials, and provides
further insight into the fundamental capabilities and limitations of
open-loop non-dissipative quantum control.


\emph{Control setting.---} Our target system is a single qubit whose
dephasing dynamics in the quantum regime is described by a diagonal
spin-boson Hamiltonian of the form $H=H_{S}\otimes I_B +H_{SB}+I_S
\otimes H_{B}$, with $H_S = \omega S_z$ and
$$ H_{SB}=S_z \otimes\sum_{k}(g_{k} b_{k}+ g_k^\ast
b_{k}^{\dagger}),\quad H_B= \sum_k\omega_{k}b_{k}^{\dagger}b_{k}.$$
\noindent
Here, $I_{S(B)}$ denote the identity operator on the system (bath),
$S_z=\hbar\sigma_z/2$ is the spin operator along the quantization
axis, and $b_{k}$ ($b_{k}^\dagger$) are canonical ladder operators for
the $k$th bosonic mode, characterized by a frequency $\omega_k$ and
coupling strength $g_{k}$.  If the bath is initially at thermal
equilibrium at temperature $1/(k_{B}\beta)$, its influence on the
qubit dynamics is encapsulated by the spectral density function
$I(\omega)\equiv\!\!\sum_{k}\vert
g_{k}\vert^{2}\delta(\omega-\omega_{k})$.  Without loss of generality,
we shall assume that $I(\omega)$
decays to zero beyond a finite ultraviolet cutoff $\omega_{c}$.

DD over an evolution interval $[0,T]$ is achieved by applying a train
of $n$ instantaneous $\pi$ pulses (each implementing a Pauli
$\sigma_x$ operator) at times $\{t_j\}$, where $0<t_1 <\ldots < t_n
<T$, and we also let $t_{0}\equiv 0$ and $t_{n+1}\equiv T$.  While
keeping the number of pulses $n$ to a minimum may be desirable for
various practical reasons, neither $n$ nor the resulting sequence
duration need to be constrained {\em a priori}.  An arbitrary long
duration $T$ may, in fact, be needed for quantum memory. In contrast,
infinite pulse repetition rates are both fundamentally impossible and
undesirable as long as $T>0$.  Let the \emph{minimum switching time}
$\tau_{\text{min}} >0$ lower-bound the smallest control time scale
achievable by any sequence:
\begin{equation}
\tau \equiv \min (t_{j+1}-t_{j})\ge\tau_{\text{min}}, \;\;\;\; j \in
\{0,\ldots, n\}.
\label{eq:taumin}
\end{equation}

If the system is initially in a nontrivial coherent superposition of
$S_z$ eigenstates, its purity in the presence of DD decays with a
factor of $\exp(-2\chi_{\{t_{j}\}})$, where the {\em decoupling error}
$\chi_{\{t_{j}\}} \geq 0$ can be exactly expressed in the following
form (see {\em e.g.} Eqs. (8c) and (10) in \cite{Uhrig2007}):
\begin{eqnarray}
\chi_{\{t_{j}\}} & = & \int_{0}^{\infty}\lambda(\omega)\,\vert
f_{\{\tilde{t}_{j}\}}(\omega)\vert^{2}d\omega,\quad \;
\tilde{t}_{j}\equiv\frac{t_{j}}{\tau},\;\;
\label{eq:chiti}\\
f_{\{\tilde{t}_{j}\}}(\omega) & = &
\sum_{j=0}^{n}(-1)^{j}(e^{i \tilde{t}_{j}\omega\tau}-
e^{i \tilde{t}_{j+1}\omega\tau }),
\label{eq:fdi}
\end{eqnarray}
and the ``spectral measure'' $\lambda(\omega)\equiv
2\coth(\beta\omega/2) I(\omega)/\omega^{2} $.  In terms of the
rescaled pulse times $\tilde{t}_{j}$, Eq. (\ref{eq:taumin}) becomes
$\tilde{t}_{j+1}-\tilde{t}_{j}\ge 1$.  Physically,
Eqs. (\ref{eq:chiti})-(\ref{eq:fdi}) can also describe the purity
decay resulting from pure dephasing in the semi-classical limit, as
due to stochastic fluctuations of the qubit energy splitting and
experimentally investigated in
\cite{Biercuk2009,Uys09,SagiAlmogDavidson2010}.  In this case,
$H_{SB}\equiv 0$ and $H_S=[\omega+\xi(t)]S_z$, where $\xi(t)$ is a
Gaussian random variable with a power spectrum $S(\omega)$
\cite{cywinski08prb}.  In order to evaluate $\chi_{\{t_{j}\}}$, it
suffices to redefine $\lambda(\omega) = S(\omega)/2\pi\omega^{2}$.
The objective of DD is to minimize $\chi_{\{t_{j}\}}$. Our main
problem then directly ties to the following: Given the fundamental
constraint of Eq. (\ref{eq:taumin}), what is a lower bound on
$\chi_{\{t_{j}\}}$?


\emph{Non-perturbative performance bound.---} A lower bound on
$\chi_{\{t_{j}\}}$ can be obtained by restricting the integral in
Eq. (\ref{eq:chiti}) to a finite range $[0,\omega_c]$, with a tight
bound ensuing if $\omega_c$ coincides with the spectral cutoff in
either $I(\omega)$ or $S(\omega)$.  We separate the dependencies of
$\chi_{\{t_{j}\}}$ upon the timings and the spectral measure
$\lambda(\omega)$ by applying Cauchy's inequality to the functions
$\lambda^{1/2} |f|$ and $\lambda^{-1/2}$:
\begin{equation}
\chi_{\{t_{j}\}}\ge \frac{1}{M_{\{\lambda\}}}
\Big(\!\int_{0}^{\omega_{c}}\!\!\!\!\vert
f_{\{\tilde{t}_{j}\}}(\omega) \vert d\omega\!  \Big)^2,\
M_{\{\lambda\}}\equiv \!\!\int_{0}^{\omega_{c}}\!\!\!\!
\frac{d\omega}{\lambda(\omega)}.
\label{dderror}
\end{equation}
Thus, the integral $\int_{0}^{\omega_{c}}\vert
f_{\{\tilde{t}_{j}\}}(\omega)\vert d\omega$, which is the $L_{1}$-norm
of the ``filter function'' $f_{\{\tilde{t}_{j}\}}$ over
$[0,\omega_c]$, determines a worst-case lower bound on
$\chi_{\{t_{j}\}}$ for {\em all} spectral densities $\lambda(\omega)$
for which the integral defining $M_{\{\lambda\}}$ is finite.

Interestingly, upon letting $e^{i\omega\tau}\equiv z \in {\mathbb C}$
in Eq. (\ref{eq:fdi}), the function $f_{\{\tilde{t}_{j}\}}(\omega)$
takes the form of a complex ``polynomial'' $P_{\{\tilde{t}_{j}\}}(z)$
with non-integer exponents.  Such {\em M\"{u}ntz polynomials} have
been studied in the mathematical literature, and a plethora of results
(and conjectures) exist on their associated norm inequalities, zeroes,
and multiplicities \cite{borwein1995polynomials}.  The (now resolved)
Littlewood conjecture in harmonic analysis \cite{Nazarov96} may be
invoked, in particular, to lower-bound the $L_{1}$-norm of
$f_{\{\tilde{t}_{j}\}}$:
\begin{equation}
\chi_{\{t_{j}\}}\ge \frac{\omega_{c}^{2}}{M_{\{\lambda\}}}\, C(\log
n)^{2}, \quad \text{if }\;\omega_c \tau> 2\pi,
\label{eq:little}
\end{equation}
with $C=O(1)$.  Also note that, regardless of $\omega_c \tau$, an
upper bound follows immediately from Eq. (\ref{eq:chiti}):
$\chi_{\{t_{j}\}}\le m_{\{\lambda\}}n^{2}$, where
$m_{\{\lambda\}}\equiv \int_0^\infty \lambda(\omega)d\omega$.
Eq. (\ref{eq:little}) implies that in the ``slow-control'' regime
where $\omega_{c}\tau>2\pi$, the DD error worsens when more pulses are
applied, and coherence may be best preserved by doing nothing.  This
reinforces how sufficiently fast modulation time scales are essential
for achieving decoherence reduction, as we discuss next.

The ``fast-control'' regime ($\omega_{c}\tau<2\pi$) is implicit in
perturbative DD treatments, where the filter function
$f_{\{\tilde{t}_{j}\}}(\omega)$ is chosen to have a Taylor series that
starts at $(\omega\tau)^{m}$, so that $\chi_{\{t_{j}\}}$ remains small
for sufficiently small values of $\omega_{c}\tau$.  While this
perturbative approach has been used for designing efficient DD
schemes, it cannot lead to a lower bound on the attainable DD error in
the presence of a timing constraint.  Consider for example UDD$_n$
sequences, in which case $t_{j}=T \sin^{2}[\pi j/(2n+2)]$ for
$j=1,\cdots,n$, and $\tau \equiv t_1$. If $\tau$ is kept fixed,
increasing $n$ is only possible at the expense of lengthening the
total duration as $T(n) =\mathcal{O}[\tau n^2]$.  Irrespective of the
fact that perturbatively the error scales as
$\mathcal{O}[(\omega_{c}\tau)^{n}]$, it carries a prefactor that grows
too fast with $n$, eventually causing the perturbative description to
break down \cite{UhrigLidar2010,Hodgson2009}.

A non-perturbative lower bound may be established by directly mapping
the $L_1$-norm integral of $f_{\{\tilde{t}_j\}}$ to the size of the
corresponding M\"{u}ntz polynomial $P_{\{\tilde{t}_{j}\}}(z)$ over an
arc of the unit circle of length $\omega_c\tau$.  Theorem 2.2 in
\cite{erdelyi2010}, in conjunction with Eq. (\ref{dderror}), then
implies:
\begin{equation}
\chi_{\{t_{j}\}}\ge \frac{1}{M_{\{ \lambda\} } \tau^2 }\, c
e^{-a/(\omega_{c}\tau)}, \quad \text{if }\;\omega_c \tau< 2\pi,
\label{eq:bound}
\end{equation}
for some numeric constants $c$ and $a$ independent of $\tau$,
$\omega_{c}$, and $\{t_{j}\}$.  The bound in Eq. (\ref{eq:bound}) is
strictly positive for spectral measures of compact support.  That it
{\em cannot} be obtained by perturbative methods is manifest from the
fact that it describes an essential singularity in $\omega_c\tau$.

It is worth to further interpret Eq. (\ref{eq:bound}) in the light of
existing results.  If the control rate is identified as the key
resource that DD leverages for removing errors, a zero lower bound on
$\chi_{\{t_{j}\}}$ would allow, in principle, arbitrarily high DD
accuracy to be achieved by using sufficiently long sequences with a
{\em fixed} $\tau_{\text{min}}$ -- that is, in analogy with
fault-tolerant quantum computation \cite{PreskillRel}, with a constant
resource overhead relative to the noise-free case.  Historically, the
impossibility of reliable computation with a constant blow-up in
resources (circuit depth) was established in \cite{Dorit} in the
broader context of noisy {\em reversible} circuits, both classical and
quantum. Therefore, our results may be taken to reinforce the
fundamental limitations of purely unitary quantum error correction,
while {\em explicitly} characterizing the way in which such limiting
performance depends upon the physical parameters.


\begin{figure}[t]
\begin{center}
\includegraphics[width=0.8\columnwidth]{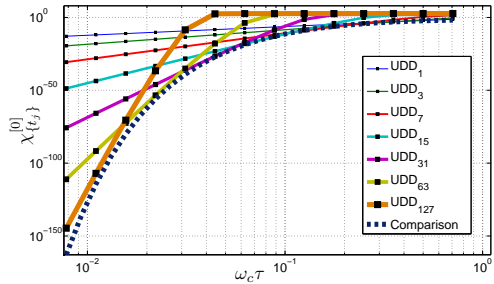}
\vspace*{-3mm}
\caption{(Color online) Decoupling error for UDD$_n$ sequences vs.
$\omega_c\tau$, for a ``flat'' spectral measure
$\lambda^{[0]}(\omega)\equiv \Theta(\omega-\omega_c)$.  The comparison
curve denotes the general lower bound, Eq. (\ref{eq:bound}), evaluated
for $a=3$, $c=1/2$, chosen to approximate a fit.
\label{fig:uddsat}}
\end{center}
\end{figure}

\emph{Achieving the performance bound.---} Note that the minimum
switching time $\tau_{\text{min}}$ enters Eq. (\ref{eq:bound})
naturally, whereas both the total duration $T$ and pulse number $n$
are markedly absent from it.  Thus interestingly, {\em if} the bound
can be achieved, it should be possible to do so irrespective of how
long $T$, provided that $n$ is unconstrained.  We can show that the
error associated with UDD sequences, $\chi^{\text{UDD}}_{n}$,
saturates the fundamental limit in Eq. (\ref{eq:bound}) in functional
form although {\em not} necessarily in {\em absolute} sense (see also
Fig. \ref{fig:uddsat}).  This follows from noting that an upper bound
to $\chi^{\text{UDD}}_{n}$ in the presence of a hard spectral cutoff
may be obtained from an upper bound to $\vert
f^{\text{UDD}}_{n}(\omega) \vert$, by tailoring $n$ to the bandwidth,
$n \equiv n_0 \approx 1/(e^2 \omega_{c}\tau)$ (see Remark 2.6 in
\cite{erdelyi2010}). This yields:
$$ \chi^{\text{UDD}}_{n}\le m_{\{\lambda\}} \hspace*{-0.5mm} \cdot
\max_{\omega \in [0, \omega_c]} \vert f^{\text{UDD}}_{n_0}(\omega)
\vert^2 \leq \frac{m_{\{\lambda\}}}{\omega_c \tau}
c'e^{-a'/(\omega_{c}\tau)}, $$
\noindent
where $c'=2/(\pi e^2),\, a'={2}/{e^2}$, and a similar functional form
as in Eq. (\ref{eq:bound}) is manifest.  With $\tau\equiv t_1 \equiv
\tau_{\text{min}}$ fixed, the duration $T$ of the ``tailored UDD$_n$''
sequences scales as
$\mathcal{O}[1/(\omega_{c}^{2}\tau_{\text{min}})]$, and the longest
allowed $\tau$-value that results in coherence improvement scales as
$1/(n\omega_c)$. Thus, UDD provides no guarantee that the error
reaches its absolute minimum and accessing the required $\tau$ becomes
increasingly harder as $T$ grows.  This motivates searching for DD
sequences that can operate {\em beyond} the perturbative regime and
retain their efficacy over the broadest range possible,
up to $1/(n\omega_c)\lesssim\tau\lesssim1/\omega_c$.

Various optimized DD strategies have been investigated for the
qubit-dephasing setting under consideration.  In ``locally optimized''
(LO) DD \cite{Biercuk2009}, optimal pulse timings are determined via
direct minimization of the error $\chi_{\{t_{j}\}}$ for a fixed target
storage time $T$, whereas in ``optimized noise filtration'' (OF) DD,
only the integral of the filter function is minimized \cite{Uys09}
(see also \cite{noise_adapted} for a noise-adapted perturbative
approach).  While LODD/OFDD can access regimes where perturbative
approaches are not efficient, they focus on matching the total
sequence duration $T$ as the fundamental constraint.  However, this
may fail to produce a satisfactory control solution if the timing
constraint imposed by Eq. (\ref{eq:taumin}) is significant.

\begin{figure}[t]
\begin{center}
\includegraphics[width=0.85\columnwidth]{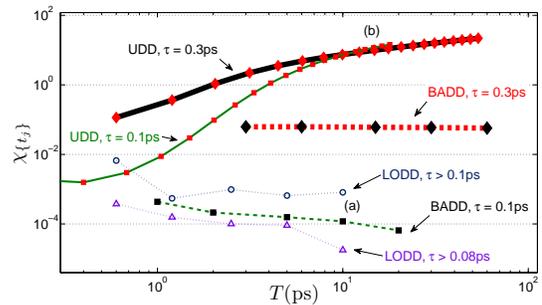}
\caption{(Color online) Decoupling error for BADD (dashed), LODD
(dotted), and UDD (solid) sequences vs. total duration $T$ with the
minimum interval $\tau$'s indicated, for a dephasing exciton qubit
operating at temperature $77$K (see text).  The search space for BADD
and LODD covers up to $n=100$ pulses, whereas for UDD $n \leq 20$.
See the Appendix for further detail and a comparison between different
pulse timing patterns.
\label{fig:badd}}
\end{center}
\end{figure}

To guarantee that such a fundamental limitation is obeyed, we
introduce optimized {\em bandwidth-adapted DD} (BADD) sequences where
{\em both} the minimum switching time {\em and} the total time are
constrained from the outset, see the Appendix for additional technical detail.  We demonstrate the usefulness of
BADD by focusing on the exciton qubit analyzed in \cite{Hodgson2009},
for which a spin-boson dephasing model with a supra-Ohmic spectral
density and a Gaussian cutoff is appropriate,
$I^{\omega_c,s}(\omega)=\alpha \omega^s \exp(-\omega^2/\omega_c^2)$
with $s=3$, $\alpha \approx 1.14\times10^{-26}$s$^{2}$,
$\omega_c\approx 3\,\text{rad}\,\text{ps}^{-1}$, and the need to avoid
unwanted excitation of higher-energy levels enforces a timing
constraint $\tau_\text{min}\approx 0.1$ ps \cite{pulsew}.  The results
are summarized in Fig. \ref{fig:badd}.  Besides indicating the
inadequacy of perturbative UDD for $T \gtrsim 1$ ps, two main features
emerge.  First, as predicted by Eq. (\ref{eq:bound}), the minimum
error achievable by BADD is mainly dictated by $\tau$, largely
independently of the total time $T$. Second, LODD performance is
fairly sensitive to the timing constraint: for a fixed $T$ ($10$ ps in
Fig. \ref{fig:badd}), ``softening'' the constraint selects LODD
sequences that outperform BADD, the opposite behavior being seen if
the constraint on the intended $\tau$ is ''hardened''.  Thus, a BADD
protocol effectively optimizes over a set of LODD sequences where the
timing constraint is only approximately met, consistent with
intuition.

\begin{figure}[t]
\begin{center}
\includegraphics[width=0.8\columnwidth]{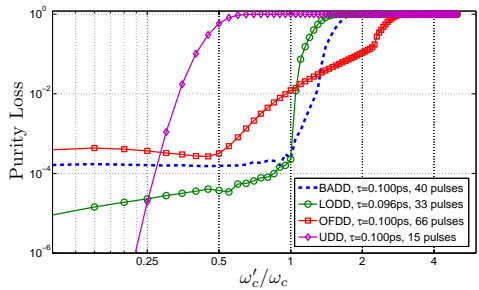}
\vspace*{-3mm}
\caption{(Color online) Purity loss, $1-e^{-2\chi_{\{t_j\} }}$,
vs. actual over presumed cutoff, $\omega_{c}^{\prime}/\omega_{c}$, for
the supra-Ohmic Gaussian spectral density ($s=3$) corresponding to the
exciton qubit. All sequences are adapted to $T\approx10$ps,
$\tau\approx0.1$ps.  Varying the ``actual'' power law of the noise to
$s=4$ and $s=2$ resulted in a qualitatively similar behavior (data not
shown).
\label{fig:Effect-of-varying}}
\end{center}
\end{figure}

In practice, an important question is whether the performance of a DD
scheme is robust against uncertainties in the underlying spectral
measure: in particular, sequences adapted to a presumed
$\omega_{c}\tau_{\text{min}}$ need not be adequate for the actual
$\omega_{c}^{\prime}\tau_{\text{min}}$.  Some illustrative results are
depicted in Fig. \ref{fig:Effect-of-varying} for sequences subject to
the same timing constraint, but applied to a setting where
$\omega_c'\ne \omega_c$.  Clearly, a smaller cutoff
$\omega_{c}^{\prime}$ leads to smaller decoherence, but much more so
for perturbative UDD sequences.
Expectedly, the knowledge of the spectral density explicitly assumed
in generating BADD and LODD results in far better coherence compared
to OFDD and UDD, especially when this knowledge is precise
($\omega_{c}^{\prime}/\omega_{c}=1$) or overestimates the
cutoff. Comparatively, BADD sequences appear to be more robust than
LODD sequences when the cutoff is underestimated.

\emph{Discussion.--- } Our mathematical description has relied on the
solvability of the dephasing spin-boson model in the limit of
instantaneous control pulses, however we expect similar timing-induced
lower bounds to exist under more general conditions.  In principle,
non-Gaussian classical dephasing such as random telegraph noise could
be addressed based on the exact solution presented in \cite{ban2010},
whereas non-bosonic dephasing models of the form $H_{SB}+H_B \equiv
S_z \otimes B_z +B_0$,
could be tackled by matching the leading-order contributions in $B_z$
with the bosonic case studied here. Note, however, that bounded timing
resources do {\em not} prevent the DD accuracy bound to be zero in
{\em special} cases -- such as ``monochromatic'' or ``non-dynamical''
baths ($H_B=0$), for both of which the length of the arc appearing in
Eq. (\ref{eq:bound}) vanishes. Similarly, ``nilpotent'' environments,
where powers of the bath operators in $H_{SB}$ and $H_B$ vanish at
some order, allow perturbative DD schemes to achieve perfect
decoupling, as perturbation theory becomes exact.  For more
``adversarial'' environments, where $H_{SB}$ is not restricted to but
includes single-axis decoherence, similar lower bounds must exist by
inclusion.  Elucidating the algebraic features responsible for a
finite vs. zero performance bound remains an interesting open problem
with implications for quantum error correction in general.  As opposed
to pulsed control scenarios, continuous-time modulation subject to
finite energy/bandwidth constraints has also been explored for
decoherence control \cite{GordonOpt}.  Although, even for a purely
dephasing qubit, finding the optimal modulation requires solving a
non-linear integro-differential equation, it would be interesting to
quantify the extent to which the extra freedom afforded by continuous
controls
may improve the achievable performance lower bounds.

We thank Michael Biercuk, Irene D'Amico, Daniel Lidar, and John
Preskill for valuable input.  Work supported from the NSF through
Grant No. PHY-0903727.

\appendix
\section*{Appendix: BADD Optimization Procedure}\label{appendix}
The BADD optimization procedure uses the following input: the spectral
measure $\lambda(\omega)$ used to characterize the environment (a
continuous positive-valued function defined on positive real numbers);
the minimum switching time $\tau$ used to constrain the pulse timings
(a positive real number); the total duration $T$ of the sequence (a
positive real number larger than $\tau$). The output from BADD is the
number of pulses $n$ (a positive integer) and the pulse intervals
$\{\tau_{j}\}_{j=1}^{n+1}$ (positive real numbers) that yield the
smallest decoupling error while satisfying the following two
constraints:
\[ (1) \; \tau_{j}\ge\tau,\quad\quad (2)\;\sum_{i=1}^{n+1}\tau_{i}=T.\]
\noindent
Both these constraints are linear in the variable \[
\mathbf{x}^{(n)}:=(\tau_{1},\tau_{2},\cdots,\tau_{n+1}).\] Also notice
that $n$ is always bounded:\[ 1\le n\le n_{\text{max}}\equiv
T/\tau-1\] The quantity to be minimized is the coherence loss over
time $T$, which is given by {[}see Eqs. (2)-(3) in the main text{]}:\[
\chi(\mathbf{x})=\int_{0}^{\infty}\lambda(\omega)\vert
f_{\mathbf{x}}(\omega)\vert^{2}d\omega,\] where
\begin{eqnarray*}
f_{\mathbf{x}}(\omega) & = &
\sum_{j=0}^{n}(-1)^{j}(e^{it_{j}\omega}-e^{it_{j+1}\omega}),\\ t_{j} &
= & \sum_{l=1}^{j}\tau_{l}.
\end{eqnarray*}

The optimization problem described by $\chi(\mathbf{x})$ is highly
non-linear and involves evaluation of $\chi(\mathbf{x})$, itself an
infinite numerical integral. Computationally, we replace the upper
limit of infinity on the integral with a parameter $\omega_{\infty}$
chosen such that
\[
\int_{0}^{\infty}\lambda(\omega)\vert
f_{\mathbf{x}}(\omega)\vert^{2}d\omega
\approx\int_{0}^{\omega_{\infty}}\lambda(\omega)\vert
f_{\mathbf{x}}(\omega)\vert^{2}d\omega.\] In the soft-cutoff exciton
qubit example considered in the main text, setting
$\omega_{\infty}=5\omega_{c}$ is more than adequate. We use the
general-purpose routines of \textsc{{Matlab}} for evaluation and
optimization of $\chi(\mathbf{x})$. In particular, we hard code the
form of $\lambda(\omega)$ into a \textsc{Matlab }function and evaluate
the integral for\textsc{ $\chi(\mathbf{x})$} using \textsc{Matlab}'s
\texttt{quadv} function (recursive adaptive Simpson quadrature).  The
numeric optimization is performed with the \texttt{fmincov} function
(linear constraints, nonlinear objective, and automatically chosen
algorithm) with a numerical tolerance of $10^{-6}$ for both the
optimal value and the constraint satisfaction. This procedure returns
an optimal choice of intervals $\mathbf{x}_{\text{min}}^{(n)}$ and a
corresponding minimum error $\chi_{\text{min}}^{(n)}$ for all possible
values of $n\in \{1,\cdots,n_{\text{max}}\}$. The minimum of
$\{\chi_{\text{min}}^{(n)}\}_{n=1}^{n_{\text{max}}}$ values identifies
the BADD optimized solution. The numerical procedure for generating
Fig. 2 in the main text takes about 4 hours to finish using 3 parallel
jobs each running on a 2.7GHz quad-core cpu.

\begin{figure}[tb]
\begin{center}
\includegraphics[width=3in]{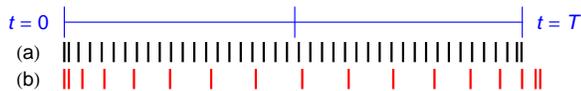}
\vspace*{-1.5mm}
\caption{(Color online) Pictorial comparison between the pulse timing
patterns for the BADD and UDD sequences corresponding to $T\approx
10$, $\tau=0.1$ps (points (a) and (b) in Fig. 2 of the main text).
Notice how the intervals of the BADD sequence (a) are compressed at
the endpoints, but become effectively uniform mid-sequence.  This
resembles the (analytical) interpolated DD protocol identified in
T. E. Hodgson, L. Viola, and I. D'Amico, Phys. Rev. A {\bf 81}, 062321
(2010).}
\end{center}
\end{figure}

In addition to the features discussed in the main text, the following
qualitative features are worth emphasizing:
\begin{itemize}

\item Theoretically, imposing constraints (1)-(2) together makes the
search space compact even including the range of possible pulse
numbers. Furthermore, Eq. (6) in the main text provides an explicit
lower bound on the objective function, independent of the duration $T$
as well as the pulse number. Both these facts provide a solid
foundation for the BADD procedure.

\item The optimization always returns $\tau_{1}=\tau$, that is, the
first interval uses the shortest available pulse interval.

\item When $\omega_{c}\tau$ is sufficiently small, the overall BADD
timing pattern approaches that of UDD pulse sequences.  Conversely,
when $\omega_{c}\tau$ is large, the timing pattern approaches a
periodic one, consisting of nearly equidistant pulses.  This is
depicted in Fig. 1 above for a BADD sequence ($n=40$) and a UDD
sequence ($n=15$) operating at $\omega_{c}\tau=0.3$, corresponding to
the smallest allowed value for the excitonic qubit discussed in the
main text.

\item Consequently, the optimal number of pulses $n$ approaches
$n_{\text{max}}$ when $\tau\omega_{c}$ is large; for smaller
$\tau\omega_{c}$ the number of pulses required is smaller than
$n_{\text{max}}$.

\item The BADD optimized error does not vary significantly with the
total time $T$, in line with the expectation that a lower bound on
decoupling error does not depend on $T$.
\end{itemize}

The LODD optimization setting mentioned in the text differs from BADD
since it uses $n$ and $T$ as the input (instead of $\tau$ and $T$),
and only constraint (2) is imposed in the minimization of the error
$\chi(\mathbf{x})$.  The OFDD optimization procedure is similar to
LODD in this respect, however it uses a specific {}``flat'' spectral
measure in which $\lambda(\omega)$ is equal to $1$ for
$\omega<\omega_{c}$ and 0 otherwise.  That is, OFDD minimizes the
integral of the filter function independently of the actual spectral
measure $\lambda$, leading to a minimum error which always
upper-bounds the one from LODD.  As also discussed in the main text,
for a given value of $T$ the pulse sequences from BADD and LODD can
effectively be matched by comparing them at (almost) equal
$\tau$'s. BADD sequences can thus be approximated by searching LODD
sequences with different pulse numbers for satisfying the minimum
interval condition (1).

\end{document}